\documentclass[preprint,showpacs,preprintnumbers,amsmath,amssymb]{revtex4}

\usepackage{graphicx}
\usepackage{dcolumn}
\usepackage{bm}
\usepackage{longtable}
\begin{document}

\preprint{APS/123-QED}

\title{Coherent Pair State of Pion in Constituent Quark Model}

\author{Masanori Morishita}
 \email{morisita@ocunp.hep.osaka-cu.ac.jp}
\author{Masaki Arima}
 \email{arima@ocunp.hep.osaka-cu.ac.jp}
\affiliation{
 Department of Physics, Osaka City University, Osaka 558-8585, Japan }

\date{\today}

\begin{abstract}

A coherent state of pions is introduced to the nonrelativistic quark model.
The coherent pair approximation is employed for the pion field
in order to maintain the spin-isospin symmetry.
In this approximation the pion is localized in the momentum space,
and the vertex form factor in the pion-quark interaction is derived from
this localization.
The nucleon masses and wave functions are calculated using this model,
and our results are compared to those of the quark model
with the one pion exchange potential.
Similar result is obtained for the mass spectrum,
but there exists a clear difference in the internal structure
of nucleon resonances.

\end{abstract}


\maketitle

\section{\label{sec1} Introduction}

The nonrelativistic constituent quark model (NRCQM) is one of the
effective models of QCD.
This model describes a baryon as a bound state of the
constituent quarks in a phenomenological confinement potential.

Isgur \textit{et al.} have successfully applied the NRCQM
to the study of baryon spectra \cite{Is78,Is79}. 
They have used the one gluon exchange potential (OGEP) with
a phenomenological confinement potential in order to
consider spin-dependence in the effective quark-quark interaction.
As shown by De R\'{u}jula \textit{et al.} \cite{Ru75}, the OGEP is introduced
so as to take account of the asymptotic freedom of QCD in the effective model.

Recently, the one meson exchange potential (OMEP)
is introduced to the NRCQM instead of using the OGEP \cite{Glo96,Glo98a,Glo98b}.
In this potential mesons belong to the pseudoscalar octet,
and are closely related with the spontaneous breaking of chiral symmetry
(SB$\chi$S).
This model has also succeeded in reproducing the observed baryon
spectra \cite{PD98}.

Many theoretical works have suggested that these mesons should be
important degrees of freedom in the NRCQM.
The OMEP seems to have desirable properties with respect to
the spin-flavor dependence of the quark-quark interaction
in contrast to the OGEP \cite{Glo96}.
The dynamical role of mesons has been considered in order to deal
with the long-standing problems in the NRCQM.
For example, the mass difference between $\Lambda$(1405) and $\Lambda$(1520)
can be produced by the coupling with the ${\bar K}N$ channel \cite{Ar90}.
In a recent work, a positive parity state appears as a first excited state
in the nucleon mass spectrum when the scalar-isoscalar excitation of
a meson field is taken into account \cite{Kow01}.

In many quark models, mesons are usually treated on the basis of
the perturbative idea.
However, the perturbative method is not appropriate for mesons
interacting strongly with quarks.
As for the OMEP, only the one particle exchange is included.
The static approximation is also used for the OMEP, while the dynamical effects
of mesons such as a self energy should not be ignored \cite{Kow98}.

In order to understand the mesonic effects on the nucleon structure,
the nonperturbative property of mesons should be considered
in a effective model.
In this paper we make use of the coherent state formalism
to describe the pion field in the NRCQM,
which is one of the nonperturbative representations of the pion field.
Instead of using Glauber's coherent state \cite{Gla63},
we apply the coherent pair approximation (CPA) \cite{Bo81} to our calculation.
In this approximation the vacuum of the pion field is constructed from
indefinite numbers of scalar-isoscalar pion pairs, and has
definite quantum numbers for the orbital angular momentum and the isospin.

Ref.~\onlinecite{Ha87} has indicated that
the hedgehog ansatz is better than the CPA as an approximation for the pion
field in a ground state nucleon \cite{Go88,Al99,Fi88}.
Because the hedgehog ansatz breaks the spin-isospin symmetry,
the quantum numbers associated with this symmetry must be projected out
by semiclassical or adiabatic method.
However, this projection method has some limitation when baryon resonances are
considered, in which the excitation energy of quarks in a baryon is
on the same order of the fluctuation energy
of the pion field \cite{Ad83,Br85,Bl88}.
Note that, because the pion field in the CPA is the eigenstate of
the spin-isospin symmetry,
we do not have to rely on the projection method.
Thus we expect that the CPA is more effective for our purpose,
as far as we are interested in the qualitative estimation of
the pion contribution on baryon spectra.

We calculate the nucleon mass spectrum by taking account of the pion field
in the CPA (the CPA model).
We apply the CPA to the study of excited baryons,
while this approximation has been used to study the ground state
nucleon \cite{Go88,Al99}.
This application is considered as a good test ground to see the role of
mesons in the nucleon structure.
We employ a variational method to determine the momentum distribution of pion,
which is similar to Tomonaga's intermediate coupling
approximation applied to the pion-nucleon and the polaron
systems \cite{To47,Le53}.
We also calculate the mass spectrum by using the NRCQM with
the one pion exchange potential (the OPEP model),
and compare the result with that of ours.

This article is organized as follows.
In Sec.~\ref{sec2} our hamiltonian for the CPA model is presented,
and the nucleon state in the CPA model is constructed in Sec.~\ref{sec3}.
The calculation with the CPA model is explained in Sec.~\ref{sec4}.
A brief review of the OPEP model are given in Sec.~\ref{sec5}.
Our results obtained by the CPA model are exhibited in Sec.~\ref{sec6},
and are compared with those by the OPEP model. 
A brief summary is given in the last section.
\section{\label{sec2} Model Hamiltonian}

We describe the nucleon as a bound state of the three constituent quarks
surrounded by the coherent state of pions.
Our model hamiltonian is written as
\begin{eqnarray}
  H=H_0+H_{\pi}+H_{\pi q} ,
\label{2.0}
\end{eqnarray}
where $H_0$ describes the relative motion of the three quarks in
the center of mass (c.m.) frame,
$H_{\pi}$ the pion kinetic energy and $H_{\pi q}$ the pion-quark interaction.

We apply the nonrelativistic kinematics for the constituent
quarks, and we write $H_0$ as
\begin{eqnarray}
  H_0 = \sum_{i=1}^3 \frac{\mathbf{p}_i^2}{2m_q}-T_{g}
       +\sum_{i<j}^3 V(r_{ij}) ,
\label{2.5}
\end{eqnarray}
where $\mathbf{p}_i$ is the momentum of the $i$th quark,
$m_q$ the constituent quark mass,
$r_{ij}=|\mathbf{r}_{ij}|=|\mathbf{r}_i-\mathbf{r}_j|$
and $V(r_{ij})$ the phenomenological confinement potential.
$T_g$ is the c.m. kinetic energy of the three quarks.
The mass difference between the $u$ and $d$ quarks is not considered
in this work.
As for the confinement potential $V(r_{ij})$,
although the linear form is often used for heavy-light quark systems
\cite{Ba97}, we choose the quadratic potential for simplicity:
\begin{eqnarray}
   V(r_{ij})=\frac{1}{6}m_{q}\omega^{2}r_{ij}^2 ,
\label{2.10}
\end{eqnarray}
where the parameter $\omega$ determines the excitation energy
of this three-quark system and its spatial distribution.

The standard form of $H_{\pi}$ is employed in this model:
\begin{eqnarray}
 H_{\pi} = \int_0^{\infty}\!\! dk\ k^{2}\omega_{k} \sum_{l_{\pi}\mu\nu}
           a^{l_{\pi}\dag}_{\mu\nu}(k)a^{l_{\pi}}_{\mu\nu}(k) ,
\label{2.20}
\end{eqnarray}
where $\omega_k=\sqrt{k^2+m_{\pi}^2}$ with the pion mass $m_{\pi}$,
and $a^{l_{\pi}}_{\mu\nu}(k)$ is the isovector annihilation operator
with the orbital angular momentum ($l_{\pi},\mu$), the isospin
component $\nu$ and the momentum $k=|\mathbf{k}|$.
The annihilation operator in the spherical representation,
\begin{eqnarray}
  a^{l_{\pi}}_{\mu\nu}(k)= i^{l_{\pi}} \int\!\! d\hat{k}
                           \ Y^*_{l_{\pi}\mu}(\hat{k})
                           a_{\nu}(\mathbf{k}) ,
\label{2.25}
\end{eqnarray}
satisfies the commutation relation
\begin{eqnarray}
 \left[ a^{l_{\pi}}_{\mu\nu}(k),\ a^{l_{\pi}'\dag}_{\mu'\nu'}(k') \right]
 = \delta_{l_{\pi}l_{\pi}'}\delta_{\mu\mu'}\delta_{\nu\nu'}
   \frac{1}{k^2} \delta(k-k') .
\label{2.30}
\end{eqnarray}

The pion-quark interaction $H_{\pi q}$ takes the nonrelativistic form
of the pseudoscalar (PS) coupling:
%
\begin{eqnarray}
 H_{\pi q}&=&\sum_{j=1}^3 H_{\pi q}^j
\nonumber\\
 = \sum_{j=1}^3&&\hspace{-6mm}
   \int\! \frac{d^{3}k}{\sqrt{2\omega_{k}(2\pi)^{3}}} \frac{g}{2m_q}
   \left(i \bm{\sigma}^j\!\!\cdot\!\mathbf{k}
         \ \bm{\tau}^j\!\!\cdot\!\bm{a}(\mathbf{k})
          e^{i\mathbf{k}\cdot\mathbf{r}_j} + \text{h.c.} \right) ,
\nonumber\\
\label{2.35}
\end{eqnarray}
%
where $g$ is the pion-quark coupling constant,
and $\bm{\sigma}^j$ and $\bm{\tau}^j$ are the spin and isospin operators
for the $j$th quark, respectively.
The pseudovector coupling is also possible for the pion-quark interaction.
Since this coupling is equivalent to the PS coupling in the low energy region,
we use the PS coupling for simplicity.
\section{\label{sec3} Basis State for Nucleon}

We construct the basis state for the representation of baryons
by making a direct product of a three-quark state and a coherent state of pions.
The three-quark state is given by a proper combination of four parts: 
\begin{eqnarray}
 |\text{3q}\rangle
    = |\left[ \psi\otimes\chi\right]^{j_q}\zeta^{t_q} C\rangle ,
\label{4.5}
\end{eqnarray}
where $\psi$, $\chi$, $\phi$ and $C$
stand for the orbital, spin, flavor and color parts, respectively.
The total angular momentum is denoted by $j_q$ and the total isospin by $t_q$.

We use the eigenstate of $H_0$ for the orbital part $\psi$.
The orbital part becomes a product of two independent
harmonic oscillators when we use the Jacobi coordinates.
The energy eigenvalue of $H_0$ is
$E_0=(N_s+3)\hbar\omega$ where $N_s=0,1,2,\cdots$.

The spin part $\chi$ has either the total spin 3/2 or 1/2,
and is the irreducible representation of the spin $SU(2)$ and
the permutation group $S_3$ simultaneously.
The symmetry of the flavor part $\zeta$ is the isospin $SU(2)$
because we deal with the non-strange baryons in this work.
Thus the flavor part is constructed in the same way as the spin part. 
The color part $C$ is the singlet representation of the color $SU(3)$.
The three-quark state $|\text{3q}\rangle$ is totally antisymmetric
under the exchange of quarks.

Each $N_s\hbar\omega$ level is degenerate.
In the case of $J\!=\!T\!=\!1/2$, for example, the degeneracy is as follows.
There is only one nucleon state for the $0\hbar\omega$ level.
The first excited level $1\hbar\omega$ is twofold degenerate, i.e.
two kinds of the $p$-wave excitation.
The second excited level $2\hbar\omega$ is fourfold degenerate, i.e.
the 1-node excitation, the $d$-wave excitation and two kinds of the $p$-wave
excitation.
These three-quark states are sequentially labeled by the subscript $i$:
$|\text{3q};i\rangle$ ($i=$1,\ 2,\ 3 $\cdots$).

The pion field in the CPA is described as the coherent pair state (CPS),
which is the simultaneous eigenstate of the orbital angular momentum and
the isospin.

The CPS is first introduced by Bolsterli \cite{Bo81}, and is applied to
studying the ground state nucleon by Goeke \textit{et al.} \cite{Go88}.
Here we summarize the properties of the CPS briefly.
See Refs. \onlinecite{Bo81,Go88,Al99} for details.
Following Bolsterli's definition \cite{Bo82},
we introduce the operator $b^{\l_{\pi}}_{\mu\nu}$ instead of using
$a^{l_{\pi}}_{\mu\nu}(k)$:
\begin{eqnarray}
  b^{l_{\pi}}_{\mu\nu} = \int_0^{\infty}
                        \! dk\ k^{2}\xi(k)a^{l_{\pi}}_{\mu\nu}(k) ,
\label{4.10}
\end{eqnarray}
with the momentum distribution function $\xi(k)$.
The function $\xi(k)$ is real, and it is normalized as
\begin{eqnarray}
  \int_0^{\infty}\! dk\ k^2\xi(k)^2 = 1 .
\label{4.15}
\end{eqnarray}
We determine $\xi(k)$ variationally in the following calculation.

The ground state of the CPS $|\text{c.p.0};\xi\rangle$, i.e. the 0-unpaired
pion state, is defined as
\begin{eqnarray}
 |\text{c.p.0};\xi\rangle =\sum_{n=0}\frac{f_{n}(x)}{(2n)!}
                    (b^{l_{\pi}\dag}\cdot b^{l_{\pi}\dag})^n|0\rangle ,
\label{4.20}
\end{eqnarray}
satisfying the eigenvalue equation
%
\begin{eqnarray}
  b^{l_{\pi}}\!\cdot\! b^{l_{\pi}}\ |\text{c.p.0};\xi\rangle
   &=&\sum_{\mu,\nu} (-1)^{\mu}
   \ b^{l_{\pi}}_{\mu\nu}\cdot b^{l_{\pi}}_{-\mu,-\nu}\ |\text{c.p.0};\xi\rangle
\nonumber\\
   &=&x\ |\text{c.p.0};\xi\rangle ,
\label{4.25}
\end{eqnarray}
%
where $x$ is the coherence parameter,
and $b^{l_{\pi}}\cdot b^{l_{\pi}}$ is scalar-isoscalar combination
for the pion pair.
Owing to the coherent property Eq.(\ref{4.25}),
the function $f_n(x)$ satisfies the recursion relation
\begin{eqnarray}
  f_{n+1}(x) = \frac{x(2n+1)}{(2L+1)(2T+1)+2n}f_n(x) .
\label{4.30}
\end{eqnarray}
We determine $f_0(x)$ so that $|\text{c.p.0};\xi\rangle$ is normalized.
The 1-unpaired pion state of the CPS is defined as
%
\begin{eqnarray}
&&\hspace{-5mm}|\text{c.p.1};\xi;l_{\pi}\mu,\nu\rangle
\nonumber\\
&&= \frac{1}{\mathcal{N}(x)}(-)^{\mu+\nu} b^{l_{\pi}}_{\mu\nu}
    |\text{c.p.0};\xi\rangle
\nonumber\\
&&= \frac{1}{\mathcal{N}(x)} \sum_{n=0}\frac{f_{n+1}(x)}{(2n+1)!}
                 \ (b^{l_{\pi}\dag}\cdot b^{l_{\pi}\dag})^{n}
                 \ b^{l_{\pi}\dag}_{\mu\nu}|0\rangle ,
\nonumber\\
\label{4.35}
\end{eqnarray}
%
where $\mathcal{N}(x)$ is the normalization factor.

Using the three-quark state (\ref{4.5}) and
the CPS (\ref{4.20}) and (\ref{4.35}),
we can write the basis states for nucleons as
\begin{eqnarray}
\left[|\text{3q};i\rangle\otimes|\text{c.p.0};\xi^i\rangle \right]^{JT} ,
\left[|\text{3q};i\rangle\otimes|\text{c.p.1};\xi^i;l_{\pi}\rangle\right]^{JT} ,
\label{4.40}
\end{eqnarray}
where $J$ and $T$ mean the total angular momentum and the total isospin of
a baryon, respectively.
Note that we attach the label $i$ to $\xi(k)$
because we consider the dependence of the CPS on the three-quark state
$|\text{3q};i\rangle$.
The $j$th nucleon resonance is written as
%
\begin{eqnarray}
  |N_j\rangle
&=& \sum_i \left(
   \alpha_{ji}\left[|\text{3q};i\rangle\otimes|\text{c.p.0};\xi^i\rangle
              \right]^{\frac{1}{2}\frac{1}{2}}
           \right.
\nonumber\\
&&\left.
  +\beta_{ji}\left[|\text{3q};i\rangle\otimes|\text{c.p.1};\xi^i;l_{\pi}\rangle
             \right]^{\frac{1}{2}\frac{1}{2}}
                   \right) ,
\label{4.45}
\end{eqnarray}
%
where the mixing coefficients satisfy
 $\displaystyle{\sum_i}(\alpha_{ji}^2+\beta_{ji}^2)\!=\!1$.
The ground state nucleon is labeled by $j=0$.
\section{\label{sec4} Variational Method}

Here we comment on our four assumptions made in our calculation. 
The first assumption is for the model space in our calculation.
Because we are interested in the low-lying nucleon states,
we truncate the three-quark states at 2$\hbar\omega$.
The same reason will hold for our second assumption, i.e.
we neglect the higher partial waves of the pion.
We consider the pion with $l_{\pi}=1$ in the following calculation.
The third assumption is for the unpaired pion state in the CPA.
We take account of the 0- and 1-unpaired pion states defined in Eqs.(\ref{4.25})
and (\ref{4.35}).
This is sufficient to see the nonperturbative effects,
although the CPS with larger number of unpaired pions should be included
for better approximation \cite{Ha87,Go88}.
The last assumption is that the c.m. of three quarks
is always at rest when each quark interacts with the pion.
Because the c.m. correction may not be negligible for precise estimation of
nucleon masses, this problem is left for our future work.

Now we construct the nucleon states with $(J,T)\!=\!(1/2,1/2)$.
The 0-unpaired pion state $|\text{c.p.0};\xi^i\rangle$ has quantum numbers
$(l_{\pi},t_{\pi})\!=\!(0,0)$, so that only the three-quark state with
$(j_q,t_q)\!=\!(1/2,1/2)$ is combined with $|\text{c.p.0};\xi^i\rangle$.
On the other hand, because the 1-unpaired pion state
$|\text{c.p.1};\xi^i;l_{\pi}\rangle$ has quantum numbers
$(l_{\pi},t_{\pi})\!=\!(1,1)$,
not only the three-quark state with $(j_q,t_q)\!=\!(1/2,1/2)$
but also the state with $(j_q,t_q)\!=\!(3/2,3/2)$ are combined with
$|\text{c.p.1};\xi^i;l_{\pi}\rangle$.
In this case we numerically checked that other possible quantum numbers
for the three-quark states, for example, $(j_q,t_q)\!=\!(1/2,3/2)$, etc.,
can be neglected in our calculation.

In order to determine the momentum distribution functions $\xi^i(k)$ and
the coherence parameters $x_i$,
we minimize the expectation value of $H$ for the ground state nucleon
variationally.
Considering the normalization condition for $\xi^i(k)$, we take the variation
with respect to $\xi^i(k)$,
\begin{eqnarray}
  \delta \left( \langle N_0|H|N_0 \rangle 
               -\sum_i c_i \int_0^{\infty}\!\! dk\ k^2\xi^i(k)^2\right)=0 ,
\label{4.0}
\end{eqnarray}
where the Lagrange multipliers $c_i$ are introduced.
Then we obtain the explicit forms of $\xi^i(k)$ in terms of
$\alpha_{0i}$, $\beta_{0i}$, $c_i$ and $x_i$.

The numerical calculation is performed iteratively as follows.
First we prepare the initial values of $\alpha_{0i}$ and $\beta_{0i}$
for the fixed values of $x_i$ and write the explicit forms of $\xi^i(k)$.
The constants $c_i$ are determined so as to normalize $\xi^i(k)$.
Next we calculate the matrix elements of $H$ by using $\xi^i(k)$,
and reevaluate the mixing coefficients $\alpha_{0i}$ and $\beta_{0i}$
by diagonalization.
This procedure is continued until these values converge.
The values of $x_i$ are chosen to give the minimum energy of
the ground state nucleon.
\section{\label{sec5} Constituent Quark Model\protect\\ with OPEP}

In this section, we briefly review the NRCQM with the one pion exchange
potential (OPEP).
We intend to compare the result of this conventional model (the OPEP model)
with that of our CPA model.
The hamiltonian is now 
\begin{eqnarray}
  H = H_0 + \sum_{i<j}^3 H_{ij}^{\text{OPEP}} ,
\label{5.0}
\end{eqnarray}
where $H_0$ is the same as that of Eq.(\ref{2.5}).
The OPEP has the form \cite{Kow98}
%
\begin{eqnarray}
  H^{\text{OPEP}}_{ij}
&=& \frac{g^2}{4\pi}\frac{1}{12m_{q}^{2}} \bm{\tau}_{i}\!\cdot\!\bm{\tau}_{j}
    \left[ S_{\pi}(\mathbf{r}_{ij}) \bm{\sigma}_{i}\!\cdot\!\bm{\sigma}_{j}
          \begin{array}{c}\ \\ \ \end{array}
    \right.
\nonumber\\
 + &&\hspace{-5mm}
    T_{\pi}(\mathbf{r}_{ij})
    \left. \left(\frac{3\bm{\sigma}_i\!\cdot\!\mathbf{r}_{ij}
                      \ \bm{\sigma}_j\!\cdot\!\mathbf{r}_{ij}}{r_{ij}^2}
                     -\bm{\sigma}_i\!\cdot\!\bm{\sigma}_j
           \right)
    \right] .
\label{5.5}
\end{eqnarray}
%

The first term of Eq.(\ref{5.5}) is the spin-spin interaction which generates
the mass difference between $N$ and $\Delta$, for example.
The spatial part $S_{\pi}(\mathbf{r}_{ij})$ of this interaction
is explicitly written as
\begin{equation}
 S_{\pi}(\mathbf{r}_{ij})=\frac{2}{\pi}m_{\pi}^2 \int_0^{\infty}\! dq
                   \frac{q^2}{q^2+m_{\pi}^2} j_0(qr_{ij}) \tilde{U}(\mathbf{q})
                   -4\pi U(\mathbf{r}_{ij}) ,
\label{5.10}
\end{equation}
where $\tilde{U}(\mathbf{q})$ is the Fourier transformation of
$U(\mathbf{r}_{ij})$.
When the pion and the quark are considered as point particles,
$U(\mathbf{r}_{ij})=\delta^{3}(\mathbf{r}_{ij})$.
This singular function is usually regularized properly to take account
of the structures of these effective particles.
Because the spin-spin interaction generally has a large effect
on the baryon masses,
the spectrum in this model depends on the choice of
this regularization function.
For example, the first excited positive parity state moves down
if the special form is used for the regularization function \cite{Glo96}.
Here we employ the simple form \cite{Kow98}
\begin{equation}
  U(\mathbf{r}_{ij})
  = \left( \frac{\kappa}{\pi} \right)^{\frac{3}{2}}e^{-\kappa r^2_{ij}} ,
\label{5.15}
\end{equation}
where the range parameter $\kappa$ is introduced.

The second term of Eq.(\ref{5.5}) is the tensor interaction.
The spatial part is
\begin{equation}
 T_{\pi}(\mathbf{r}_{ij})=\frac{2}{\pi} \int_0^{\infty}\! dq
                 \frac{q^4}{q^2+m_{\pi}^2} j_2(qr_{ij}) \tilde{U}(q) .
\label{5.20}
\end{equation}
The tensor force generates the mass splitting between $N$(1535) and $N$(1650)
and has large effects on the internal structures of these states.

Note that in the OPEP model the nucleons are purely expressed by the
three-quark states $|\text{3q};i\rangle$,
while in the CPA model the coherent pair states are included.
We consider those states with $(j_q,t_q)=(1/2,1/2)$ up to the
$2\hbar\omega$ energy level in this model.
\section{\label{sec6} Results}

We calculate the nucleon mass spectrum and the ground state energy of $\Delta$
by using the CPA model.
The parameters are determined as follows.
The experimentally observed value is used for the pion mass, $m_{\pi}=140$ MeV.
The constituent quark mass $m_q$ is fixed at 300~MeV.
The strength $\omega$ of the confinement potential and the pion-quark coupling
constant $g$ are chosen so that the energy differences between
the ground state nucleon and the negative parity states agree with
the experimental data \cite{PD98}.

We also calculate the mass spectrum by using the OPEP model,
and the result is compared with that in the CPA model.
Common values are used for the $m_q$ and $m_{\pi}$ in these models,
and the other parameters ($\omega$, $g$ and $\kappa$) are determined
in the same way as the CPA model. 
\subsection{Energy Levels of Negative Parity States and $\Delta$}
 
Both in the CPA model and in the OPEP model,
we can find the parameter sets which reproduce the observed excitation energies
of negative parity states.
The splitting of these states is due to the tensor force generated by the
pion-quark interaction.
The numerical values of parameters are summarized in Table~\ref{tab1}.
The obtained mass spectra are displayed in Fig.~1.
\begin{figure}[h]
  \includegraphics{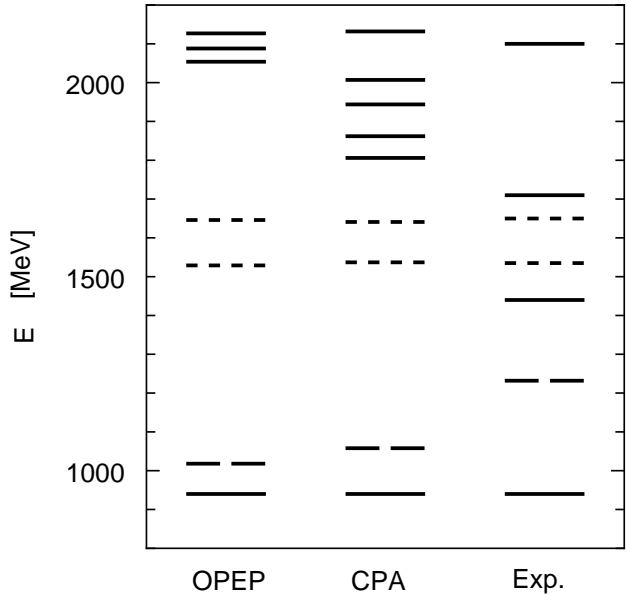}
  \caption[]{Excitation spectra of the nucleon.
            The solid and short-dashed lines show the masses of positive parity
            and negative parity states, respectively.
            The ground state energy of $\Delta$ is shown by
            the dashed line for each case.}
  \label{fig1}
\end{figure}
\begin{table}[h]
 \caption[]{Parameter sets in the CPA model and the OPEP model.}
 \label{tab1}
   \begin{ruledtabular}
    \begin{tabular}{cccc}
    & $\omega$ & $g$ & $\kappa$       \\ \hline
CPA &  380 MeV &  5.10        & $-$           \\
OPEP&  550 MeV &  1.75        & 5.0 fm$^{-2}$ \\
    \end{tabular} 
   \end{ruledtabular}
\end{table}

The $N$-$\Delta$ mass difference becomes about 120~MeV in the CPA model
and about 80~MeV in the OPEP model.
The result for this quantity is somewhat improved in the CPA model,
even though both values are smaller than the observed value $\sim$290~MeV.

Except for the values of model-dependent parameters,
the qualitatively similar results are obtained by the two different models.
As far as the pion and its interaction with quarks are considered in the NRCQM,
it is difficult to reproduce the masses of negative parity nucleons and
the $N$-$\Delta$ mass difference simultaneously.
The complete description of the nucleon spectrum is not realized
even if the nonperturbative coherent property is introduced to the pion field.

Let us discuss the difference between the two models by considering the matrix
elements of $H_{\pi q}$.
In the CPA model, the matrix element between $|\text{c.p.0};\xi^{\alpha}\rangle$
and $|\text{c.p.1};\xi^{\beta};1\mu,\nu\rangle$ becomes
\begin{eqnarray}
 &&\hspace{-5mm}\langle \text{c.p.0};\xi^{\alpha}|H_{\pi q}|
           \text{c.p.1};\xi^\beta;1\mu,\nu\rangle
\nonumber\\
 &&=\sum_j
    \int\! \frac{d^{3}k}{\sqrt{2\omega_{k}(2\pi)^{3}}}\frac{g}{2m_{q}}
       \bm{\sigma}^j\cdot\mathbf{k}
        \tau^{j\dag}_{\nu}\ Y_{1\mu}(\hat{k})
\nonumber\\
 &&\times
 \left[\xi^{\beta}(k)F^{+}(x_{\alpha},x_{\beta})e^{i\mathbf{k}\cdot\mathbf{r}_j}
   + \xi^{\alpha}(k)F^{-}(x_{\alpha},x_{\beta})e^{-i\mathbf{k}\cdot\mathbf{r}_j}
 \right] ,
\nonumber\\
\label{6.0}
\end{eqnarray}
where $F^{+}(x_{\alpha},x_{\beta})$ and $F^{-}(x_{\alpha},x_{\beta})$ are
given by
%
\begin{eqnarray}
  F^{+}(x_{\alpha},x_{\beta})
 &=&\frac{f_1(x_{\beta})}{\mathcal{N}(x_{\beta})} \sum_{n=0}^{\infty}
  \frac{f_n(x_{\alpha})}{(2n)!}x_{\beta}^n s_{\alpha\beta}^{2n}\ ,
\nonumber\\
  F^{-}(x_{\alpha},x_{\beta})
 &=&\frac{f_1(x_{\alpha})}{\mathcal{N}(x_{\beta})} \sum_{n=0}^{\infty}
  \frac{f_{n+1}(x_{\beta})}{(2n+1)!}x_{\alpha}^n
  s_{\alpha\beta}^{2n+1} ,
\label{6.2}
\end{eqnarray}
%
and $s_{\alpha\beta}$ is defined as
\begin{eqnarray}
 s_{\alpha\beta} = \int_0^{\infty}\! dk\ k^2 \xi^{\alpha}(k)\xi^{\beta}(k) .
\label{6.3}
\end{eqnarray}
We make an effective interaction which is active only for the model space
composed of the three-quark and the 0-unpaired pion states,
\begin{eqnarray}
&&\hspace{-5mm}H_{\text{eff}}^{\text{CPA}}
 = \langle \text{c.p.0};\xi^{\alpha}|H_{\pi q}|\text{c.p.1};\xi^\gamma\rangle
\nonumber\\
&&\times\frac{1}{E-\langle \text{c.p.1};\xi^{\gamma}|H_{\pi}|
                      \text{c.p.1};\xi^\gamma\rangle}
  \langle \text{c.p.1};\xi^{\gamma}|H_{\pi q}|\text{c.p.1};\xi^\beta\rangle .
\nonumber\\
\label{6.4}
\end{eqnarray}

In the OPEP model, the matrix element between $|0\rangle$
(the normal vacuum for the pion field) and 
$|\mathbf{q},\nu \rangle = a^{\dag}_{\nu}(\mathbf{q})|0\rangle$ is
\begin{equation}
  \langle 0| H_{\pi q}^j |\mathbf{q},\nu\rangle
 =\frac{1}{\sqrt{2\omega_{q}(2\pi)^{3}}}\frac{g}{2m_{q}}
    i\bm{\sigma}^j\cdot\mathbf{q} e^{i\mathbf{q}\cdot\mathbf{r}_j}
        \tau^{j\dag}_{\nu} \sqrt{\tilde{U}(q)} ,
\label{6.5}
\end{equation}
where $\mathbf{q}$ is the momentum of exchanged pion
and $\tilde{U}(q)$ the form factor introduced in Sec.~\ref{sec5}.
By using this matrix element, we can write the OPEP as
\begin{eqnarray}
 H^{\text{OPEP}}
 &=& \sum_{i<j} \left( \sum_{\nu}\int\! d^3q  
               \langle 0| H_{\pi q}^i |\mathbf{q},\nu\rangle \frac{1}{-\omega_q}
                 \langle \mathbf{q},\nu| H_{\pi q}^j |0\rangle
                 \right.                
\nonumber\\
 && \hspace{10mm}\left. \begin{array}{c}\ \\ \ \end{array}
            +\left( i \leftrightarrow j \right)
                 \right) .
\label{6.10}
\end{eqnarray}
Note that the closure is assumed for the intermediate baryons in this process.

We first comment on the vertex form factor appeared in the matrix elements
(\ref{6.0}) and (\ref{6.5}).
The vertex form factor is necessary for the low energy effective model in order
to suppress the high momentum contribution of pion.
The form factor also suppresses
the coupling with highly excited baryons and higher partial waves of
pion included implicitly in the intermediate states.
Many kinds of explanation are given for the origin of the form factor,
e.g. due to the finite size of the pion and quark.
Thus in the  OPEP model we must supply the form factor $\tilde{U}(q)$
to the pion-quark interaction with an additional parameter $\kappa$.

\begin{figure}[] 
 \includegraphics{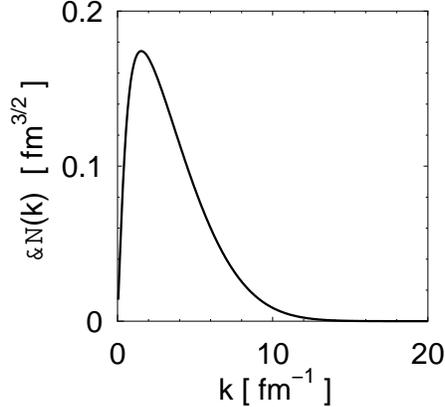}
  \caption[]{Momentum distribution function $\xi^0(k)$ of the pion field
             for the three-quark ground state with $(j_q,t_q)=(1/2,1/2)$.}
 \label{fig2}
\end{figure}

In the CPA model, we consider only the $p$-wave pion.
The momentum distribution function $\xi^{\alpha}(k)$
in the matrix element Eq.(\ref{6.0}), which is displayed in Figure~\ref{fig2},
plays the role of a vertex form factor
like $\tilde{U}(q)$ in the OPEP model.
The coherent property accounts for the origin of the form factor:
the coherent state of pions is localized,
and high momentum contribution of pion is cut off.
The form factor in the CPA model does not include any free parameters,
and is self-consistently determined by the variation.

Secondly, we notice that the values of $g$ in the two models are
different.
In the OPEP model all the continuum states of the plane wave pion
(except for the high momentum components) contribute to the interaction.
On the other hand, in the CPA model the pion in the coherent pair state
always has the localized distribution specified by $\xi^{i}(k)$,
and continuum states of pion is excluded from this model.
Because of this constraint on the pion state,
the matrix elements of $H_{\text{eff}}^{\text{CPA}}$
become smaller than those of $H^{\text{OPEP}}$.
The large value of $g$ is required in the CPA model in order to
reproduce
the observed spectrum, while the smaller value of $g$ is found
in the OPEP model.

Here we comment on our assumption about the partial wave of pion in
the intermediate state.
As demonstrated by our calculation,
we obtain the qualitatively similar results
in the two models for the masses of negative parity nucleons and
the $N$-$\Delta$ mass difference.
The use of the large value of $g$ compensates for neglecting
all partial waves except the $p$-wave in the CPA model.
We conclude that the contribution to the nucleon masses is dominated by
the $p$-wave pion in the NRCQM including the pionic degree of freedom.
\subsection{Mixings Coefficients}

Table~\ref{tab2} shows the mixing coefficients for the negative parity excited
states.
In both models these two resonances are mostly due to the $p$-wave excitation
of quarks in the nucleon, and are also mixtures of the two basis states
with spin 1/2 and 3/2 for the three-quark part.
The structure of these states in the CPA model is found to be
different from that in the OPEP model.
\begin{table*}[h]
 \caption[]{Mixing coefficients of the negative parity states.
      Masses Calculated for each state are displayed in the second column.
      Both states include the spatial part with the $p$-wave excitation
      of one quark (thus the parity is negative),
      and the flavor part with mixed symmetry ($t_q=1/2$).
      These states are distinguished by the total intrinsic spin $s_q$
      ($s_q=1/2,3/2$).
      Both states have the total spin $j_q=1/2$.}
 \label{tab2}
 \begin{ruledtabular}
  \begin{tabular}{ccccc}
 CPA\ &mass[MeV]\ &\ $[|\text{3q};s_q=3/2\rangle\otimes|\text{c.p.0}\rangle]
                      ^{\frac{1}{2}\frac{1}{2}}$
                \ &\ $[|\text{3q};s_q=1/2\rangle\otimes|\text{c.p.0}\rangle]
                      ^{\frac{1}{2}\frac{1}{2}}$
                \ &\ $\cdots$\\
\hline
    &1537& 0.682      & 0.652    & $\cdots$  \\
    &1641& 0.690      & $-$0.720 & $\cdots$  \\
\hline \hline
 OPEP\ &mass[MeV]\ & $|\text{3q};s_q=3/2\rangle$ \
                   & $|\text{3q};s_q=1/2\rangle$
                   &
\\ \hline
    &1529& 0.399      & 0.917    & \\
    &1646& 0.917      & $-$0.399 & \\
  \end{tabular}
 \end{ruledtabular}
\end{table*}

Let us consider the mixture of the two basis states,
which is generated by the tensor force appearing in $H^{\text{OPEP}}$ and
in the effective interaction $H^{\text{CPA}}_{\text{eff}}$.
There are three matrix elements of the hamiltonian $H$ with respect to
these basis states:
two diagonal elements $\langle H\rangle_{1/2}$, $\langle H\rangle_{3/2}$ and
an off-diagonal element $\langle H\rangle_{1/2,3/2}$.
The mixing coefficients are solely determined by the ratio
$(\langle H\rangle_{3/2}-\langle H\rangle_{1/2})$ : 
$\langle H\rangle_{1/2,3/2}$.
Calculating the matrix elements,
we obtain 0.1~:~1 when we use the hamiltonian
$H=H_{\pi}+H^{\text{CPA}}_{\text{eff}}$ in the CPA model,
and 1.9~:~1 in the OPEP model with $H=H^{\text{OPEP}}$.
The difference in this ratio between the two models can be explained
if we consider the intermediate baryons included in the effective interactions
(\ref{6.4}) and (\ref{6.10}).
The mixing coefficients of negative parity nucleons are sensitive
to the model space for baryons considered in the effective model.

The magnitudes and the relative sign of the mixing coefficients manifest
the internal structures of $N$(1535) and $N$(1650).
It is well known that the configuration mixing in the OPEP model
is not appropriate to the analyses
of $\pi N$ and $\eta N$ reactions \cite{Ar92}.
The CPA model does not improve this shortcoming of the OPEP model.
As explained in Ref. \onlinecite{Yo00},
the meson-quark coupling is not sufficient
to explain these reactions, and we need something else for the quark-quark
interaction.

The structures of $N$(940) and $\Delta$(1232) are shown in Tables~\ref{tab3}
and \ref{tab4}, respectively.
In both models, $N$(940) is dominated by the three-quark state
in the lowest energy. 
This is also the case for $\Delta$(1232), and the $d$-wave excitation of quarks
is hardly found in this state \cite{Wo92,No90,Ta85}.
We note that $N$(940) and $\Delta$(1232) in the CPA model have the three-quark
component with $(j_q,t_q)=(3/2,3/2)$ and $(j_q,t_q)=(1/2,1/2)$, respectively.
This is because the CPA model includes the 1-unpaired pion state
in the description of baryons.
\begin{table*}[h]
 \caption[]{Mixing coefficients of $N$(940).
   The quarks are always in the $0\hbar\omega$ state in this table.
   $|\text{3q};N\rangle$ stands for the three-quark state with
   $(j_q,t_q)=(1/2,1/2)$,
   and $|\text{3q};\Delta\rangle$ with $(j_q,t_q)=(3/2,3/2)$.}
 \label{tab3}
 \begin{ruledtabular}
  \begin{tabular}{ccccc}
CPA &$[|\text{3q};N\rangle\otimes|\text{c.p.0}\rangle]^{\frac{1}{2}\frac{1}{2}}$
\ &\ $[|\text{3q};N\rangle\otimes|\text{c.p.1}\rangle]^{\frac{1}{2}\frac{1}{2}}$
\ &\ $[|\text{3q};\Delta\rangle\otimes|\text{c.p.1}\rangle]
       ^{\frac{1}{2}\frac{1}{2}}$
     & $\cdots$
\\ \hline
     & 0.860 & $-$0.293 & $-$0.334 & $\cdots$
\\  \hline\hline
 OPEP& $|\text{3q};N\rangle$
     & $\cdots$
     &
     &
\\ \hline
     & 0.998  & $\cdots$     &     &    
\\
  \end{tabular}
 \end{ruledtabular}
\end{table*}
\begin{table}[h]
 \caption[]{Mixing coefficients of $\Delta$(1232).
            The notation is the same as in Table~\ref{tab3}.}
 \label{tab4}
 \begin{ruledtabular}
  \begin{tabular}{cccc}
CPA & $[|\text{3q};\Delta\rangle\otimes|\text{c.p.0}\rangle]
       ^{\frac{3}{2}\frac{3}{2}}$
\ &\ $[|\text{3q};N\rangle\otimes|\text{c.p.1}\rangle]^{\frac{3}{2}\frac{3}{2}}$
\ &\ $\cdots$
\\ \hline
     & 0.898 & $-$0.389 & $\cdots$ 
\\  \hline\hline
 OPEP & $|\text{3q};\Delta\rangle$
      & $\cdots$
      &
\\ \hline
      & 0.9998 & $\cdots$     &
\\
  \end{tabular}
 \end{ruledtabular}
\end{table}
\subsection{First Excited State of Positive Parity Nucleon}

The first excited state of the nucleon is called the Roper resonance $N(1440)$.
The excitation energy of this resonance is hard to reproduce in the constituent
quark model without introducing strong short-range attractive force
among the quarks \cite{Glo96,Glo98a,Glo98b}.
Several recent works have suggested that the positive parity state can be
interpreted as a scalar-isoscalar excitation of some extra degrees of
freedom \cite{Kow01,Mo92}.
The first positive parity excitation in our calculation does not correspond to
the observed Roper resonance but to another state with higher mass.
The coherent state of pions in the CPA does not produce such a
strong short-range force as suggested by Refs.
\onlinecite{Glo96,Glo98a,Glo98b}.
As for the spectrum of positive parity nucleons, the CPA model is similar
to the OPEP model.

However, the internal structure for the positive parity excitation is
remarkably different between the two models.
In the OPEP model, the first positive parity excitation is mainly due to
the nodal excitation of quarks (about 97$\%$)
which has the symmetric spatial part.
On the other hand, in the CPA model,
the components of this excited state are classified as follows:
the three quarks in the lowest energy with the 1-unpaired pion (about 47$\%$),
and the nodal excitation of quarks with the 0-unpaired pion (about 38$\%$).
This nodal excitation has the mixed symmetric spatial part
in contrast to the symmetric excitation in the OPEP model.

This result shows that the three-quark state accompanied by
the 1-unpaired pion constitutes an additional positive parity state
in the nucleon spectrum.
This is the new aspect of our study taking account of the explicit role of
pion in the NRCQM.
However, there is no Roper-like excitation below the negative parity nucleons
because the additional state (pion mode excitation in the nucleon) has
an excitation energy on the order of 2$\hbar\omega$. 
\section{Summary}

We have introduced the coherent state of pions to the NRCQM in the CPA.
The coherent property of the pion does not change the prediction of the OPEP
model drastically.
As for the nucleon mass spectrum, the CPA model gives qualitatively similar
results in comparison with the OPEP model.
However, the difference between the two models is clearly seen
in the pion-quark interaction and the internal structure of baryons.

In view of the SB$\chi$S,
the pion is also an important constituent of the nucleon
as well as the constituent quarks.
Because the mass spectrum generated only by the excitation of quarks is not
satisfactory for explaining the observed nucleon spectrum,
the excitation of extra degrees of freedom, such as the pion in our work,
should be considered in the analyses of the nucleon mass spectrum.
\begin{acknowledgments}
The authors thank Prof.~Y.~Sakuragi and
the nuclear theory group of Osaka-City University
for useful discussions.
\end{acknowledgments}
\bibliography{ref}
%
%

\end{document}